\documentclass[%
 reprint,
 amsmath,amssymb,
 aps,
]{revtex4-2}

\usepackage{graphicx}% Include figure files
\usepackage{dcolumn}% Align table columns on decimal point
\usepackage{bm}% bold math
\begin{document}

\preprint{APS/123-QED}

\title{Annihilation energy and decay time of an ortho-Positronium}

\author{Abdullah Guvendi}
  \email{abdullahguvendi@gmail.com}
 \altaffiliation[Also at ]{Department of Physics, Science Faculty, Akdeniz University, TR- 07058
Antalya, Turkey.}%Lines break automatically or can be forced with \\
\author{Yusuf Sucu}%
 \email{yusufsucu@akdeniz.edu.tr}
\affiliation{%
 Department of Physics, Science Faculty, Akdeniz University, TR- 07058
Antalya, Turkey}%

\date{\today}% It is always \today, today,

\begin{abstract}
We approach to the ortho-positronium (o-Ps) as a relativistic two-body problem in $2+1$ dimensions in which o-Ps is composed of two-oppositely charged particles interacting via an attractive Coulomb force. In addition to separation of center of mass and relative coordinates, mapping the background into the polar space-time gives possibility of construction of possible spin eigen-states of o-Ps. This approach makes the energy spectrum complex in order to describe o-Ps that can decay. From the complex energy expression, we find the annihilation energy, binding energy and the life-time of o-Ps, in S state.

\end{abstract}

\keywords{positronium, annihilation energy, life-time of ortho-positronium, triplet-spin state, relativistic two-body problem, nuclear medicine.}%Use showkeys class option if keyword
\maketitle

%\tableofcontents

\section{\label{sec:level1}INTRODUCTION}
The existence of an anti-particle having equal mass with an electron and carrying an opposite charge was postulated by Dirac in 1928 \cite{dirac1}. Experimental observation of these particles called as positron \cite{anderson} was made by C. Anderson in cosmic ray research in 1932 and then called as positronium (Ps) which is a bound state of positron and electron experimentally discovered by Deutsch in 1951 \cite{4}. The Ps is a unique system which is lightest hydrogen-like and purely leptonic atom which is not affected by the finite size effects. Due to the leptonic atom bound by an electromagnetic potential it is not only an eigen-state of parity operator, but also similarly to the flavor neutral mesons because of the its symmetry under the ex-change of particles, it is also eigen-state of the charge conjugation operator.

The Ps, in ground state with orbital angular momentum $l=0$, can be formed in a triplet ($1^{3}S_{1}$, parallel spin orientation) and singlet ($1^{1}S_{0}$, anti-parallel spin orientation) state. The both ground states of Ps, $1^{3}S_{1}$ and $1^{1}S_{0}$, are known as ortho-positronium (o-Ps) and para-positronium (p-Ps), respectively. Due to the being a bound state of a particle and its antiparticle counterpart, Ps is a intrinsically meta-stable atom and indicates a self-annihilation with emitting even or odd number gamma quantas due to the C-symmetry. The annihilation process requires the overlapping of the both particle wave functions at the origin of center of mass frame \cite{2,3}. Even though higher order annihilation is possible \cite{4a}, the annihilation process occurs predominantly $l=0$. Probabilities of the production of Ps in both ground states are $1/4$ for p-Ps and $3/4$ \cite{b} for o-Ps without any manipulation such as an external field \cite{a}. Due to the odd-parity under C-transformation \cite{h}, o-Ps decays into an odd gamma quanta on the other hand the p-Ps decays an even gamma quanta. They decay predominantly into three and two gamma quantas, respectively, because of smallness and kinematics of the fine structure constant, where charge conjugation of the photons (it is actually odd) defines the number of gamma quanta. Furthermore, especially for the $n=2$ ($n$ is principle quantum number) situation, excitation of the system to $l>0$ is the fundamental aim to manipulate the annihilation rate via applying an external electromagnetic field so that the life-time of the Ps can be enhanced \cite{12}.

There are experimental results and theoretical calculations for decay time of a positronium in the literature (these decay times vary in the range of nanoseconds and picoseconds) \cite{a1,b1}. The life-time differences resulting from phase-space and additional $\alpha$ suppression factor have been used for the observation of unknown interaction which is not accommodated in the Standard Model (SM) \cite{kk}, particularly in o-Ps decay process.

Since the o-Ps is an eigenstate of charge-parity (CP) operator combination \cite{a}, it might allow for the investigation of the discrete symmetries in the purely leptonic sector \cite{b}. Furthermore, it is thought that Ps can be used for detection of the quantum gravity effect \cite{h,6} of an anti-electron because it is also an anti-atom. Over a few decade, in the hadronic sector, CP-violation has been demonstrated \cite{o}, but has not been observed in the leptonic sector \cite{l}, already. Nowadays, since the mentioned symmetry violations are very small for the observed baryons \cite{m,n}, researchers have been trying to understand the origin of the matter-antimatter asymmetry in the universe especially in purely leptonic sector and therefore it requires a new source to detect the violations or a new mechanism. So, the purely leptonic atom may be a good candidate and the violating effect may be understood in annihilation process of o-Ps \cite{p}, if CP-symmetry is not conserved. Because of the photon-photon interaction, the prediction of the SM for the violence is approximately $10^{-10}$ \cite{l}, in the final state, but this value is much small to be observed. Furthermore, this decay also known as invisible or rare decay of the o-Ps attracts much attention by the research community due to the existence of the mirror matter pair could be candidate of dark matter \cite{r}. The discovery of decay time or annihilation energy of Ps contributes to radioactivity, Einstein's special relativity and laws of conservation of the physical properties and it helps for revolution of medical imaging technology via positron emission tomography (PET) \cite{v} devices in which the working principle of these devices depends on the self-annihilation character of the Ps. Standard PET scanners in which inorganic crystals are used employ the electron-positron annihilation into two gamma quanta and the PET devices have been used in routine imaging technology for specific biological processes occurring in the body such as neurology \cite{y} and oncology \cite{z}, over a few decades. Recently, in Cracow, Jagiellonian-positron emission tomography (J-PET) device is the announced operating with organic scintillators as opposed to the standard PET devices \cite{2} and based on signature of three gamma quantas \cite{b} causing from an o-Ps annihilation. This device can support the observation of discrete symmetries in decays of o-Ps besides the medical monitoring.

The information carried by $3^{rd}$ gamma quanta allows us to define the position of annihilation point, although the radiations arising from a positron-electron pair annihilation occurs outside of the positron emitter source in J-PET. This multi-purpose device carries a potential to become a newest technology to discover some ideas in new physics \cite{h} researches, particularly in particle physics. Therefore, acquiring exact and correct energies of gamma quantas resulting from o-Ps annihilation besides their place on the detector provide precise definition of localization point of the tissue or tumor in which the annihilation occurs. Moreover, detection of the three quantas with a better sensitivity may give also information about the matter background \cite{b} because the annihilation process appears in the localization point of the tumor. Therefore, we can obtain crucial information about the tumors or tissues such as their physical properties by using J-PET devices.

All of the reasons listed above, here, we present an exact solution of the annihilation energy, binding energy and decay time of o-Ps as a relativistic two-body interacting via an attractive Coulomb force. The calculation procedure is explained as folows. First, by writing the fully covariant two-body equation in $(2+1)$- dimensional spacetime background and then by separating to the relative and center of mass variables, we obtained a first order coupled differential equation set provided that the center of mass does not carry momentum. Mapping the background into the polar space-time so that we can construct to possible spin eigen-states of o-Ps $m_{s}$, $(+1,0,-1)$ yield possibility of deriving a second order radial differential wave equation for a central interaction energy term by $m_{1}=m_{2}$. Then, the spectrum given in single equation (see Eq.(\ref{fre})) is modified for electron-positron pair and we found the exact annihilation energy and life-time of o-Ps in S-state simultaneously together which differs our current work with the existing literature.

%\subsection{\label{sec:level2}Second-level heading: Formatting}

\section{\label{a1} Two-body Equation in (2+1) Dimensions}

The fully covariant two-body equation \cite{barut,5} in $(2+1)$-dimensional spacetimes for interacting two particles is written as in the following;
\begin{eqnarray}
\begin{split}
&\left[ \left( \sigma _{\mu }^{\left( 1\right) }i\hslash \left( \partial
_{\mu }^{\left( 1\right) }+\frac{ie_{1}}{\hbar c}A_{\mu }^{\left( 2\right) }\right)
-m_{1}cI\right) \otimes \sigma _{0}^{\left( 2\right) }\right.\\
&\left.+\sigma _{0}^{\left( 1\right) }\otimes \left( \sigma _{\mu }^{\left( 2\right) }i\hslash \left(
\partial _{\mu }^{\left( 2\right) }+\frac{ie_{2}}{\hbar c}A_{\mu }^{\left( 1\right) }\right)
-m_{2}cI\right) \right] \phi \left( \mathbf{x}_{1},\mathbf{x}_{2}\right) =0. \label{Eq1}
\end{split}
\end{eqnarray}
Here, bispinor function is given as follows.
\begin{eqnarray}
&\phi\left( \mathbf{x}_{1},\mathbf{x}_{2}\right) =\Phi \left( \mathbf{x}_{1}\right) \otimes \Omega \left( \mathbf{x}_{2}\right)\nonumber\\
&\Phi \left( \mathbf{x}_{1}\right) =\binom{\varphi \left( \mathbf{x}_{1}\right) }{\varkappa \left( \mathbf{x}_{1}\right) },\Omega \left( \mathbf{x}_{2}\right) =\binom{\varphi \left( \mathbf{x}
_{2}\right) }{\varkappa \left( \mathbf{x}_{2}\right) }\label{Eq00}
\end{eqnarray}
Besides $\otimes$ symbols indicate Kronocker production and $
e_{1}, e_{2}, m_{1}, m_{2}$ are charges and masses of the first and second particles, respectively. In addition to the unit matrice (I), general position vectors $x_{1}$ and $x_{2}$, $\sigma _{\mu }^{\left( 1,2\right) }\left(
x_{1},x_{2}\right)$ \ \ are the space-depended Dirac matrices. The well-known center of mass and relative variables are introduced as follows \cite{5}.
\begin{eqnarray}
&&\mathbf{R}=\frac{1}{M}\left( m_{1}\mathbf{X}_{1}+m_{2}\mathbf{X}_{2}\right)\nonumber\\
&&\mathbf{r}=\mathbf{X}_{1}-\mathbf{X}_{2}\nonumber\\
&&M=m_{1}+m_{2}\nonumber\\
&&\mathbf{X}_{1}=\left( \overset{\left( 1\right) }{t},\overset{\left(
1\right) }{x},\overset{\left( 1\right) }{y}\right)\nonumber\\
&&\mathbf{X}_{2}=\left( \overset{\left( 2\right) }{t},\overset{\left(
2\right) }{x},\overset{\left( 2\right) }{y}\right)\nonumber\\
&&\overset{\left( 1\right) }{\partial _{X_{\mu }}}=\partial _{r_{\mu }}+\frac{%
m_{1}}{M}\partial _{R_{\mu }}\nonumber\\
&&\overset{\left( 2\right) }{\partial _{X_{\mu }}}=-\partial _{r_{\mu }}+%
\frac{m_{2}}{M}\partial _{R_{\mu }}\nonumber\\
&&\overset{\left( 1\right) }{\partial _{X_{\mu }}}+\overset{\left( 2\right) }{%
\partial _{X_{\mu }}}=\partial _{R_{\mu }}\nonumber\\
&&\partial _{r_{\mu }}=\frac{1}{M}\left( m_{2}\overset{\left( 1\right) }{%
\partial _{X_{\mu }}}-m_{1}\overset{\left( 2\right) }{\partial _{X_{\mu }}}%
\right)\nonumber\\
&&\partial _{\mu }=\left( \frac{\partial _{t}}{c},\partial _{x},\partial
_{y}\right)\ \ \ \mu =0,1,2 \label{aaaa1}
\end{eqnarray}
In above definitions, $\mathbf{X}_{1}$ and $\mathbf{X}_{2}$ are the position vectors of the first and the second particles, respectively. It is known that, for a charge-charge interaction, $A_{0}$ is
\begin{eqnarray}
A_{0}=V\left( \mathbf{x}_{1}-\mathbf{x}
_{2}\right). \label{potential}
\end{eqnarray}
In $(2+1)$ dimensions, we can chose the matrices satisfying Dirac algebra \cite{20} in terms of constant Pauli matrices as follows.
\begin{eqnarray}
\sigma _{0}=\sigma ^{z},\quad \sigma _{1}=i\sigma
^{x},\quad \sigma _{2}=i\sigma ^{y} \label{matrices}
\end{eqnarray}
At this point, we can separate the space-time coordinates using the following definitions where each corresponds to the components of bispinor in Eq.(\ref{Eq00})
\begin{eqnarray}
&D_{p}\left( r,R,R_{0}\right):=\zeta _{p}\left( r\right) \Phi_{p} \left( R\right) e^{-iwR_{0}},\nonumber\\
&\Phi_{p}\left( R\right) :=e^{-i\mathbf{k.R}},\ \ (p=1,2,3,4). \label{components}
\end{eqnarray}
Here, $R_{0}$ and $R$ represent for time and space coordinates of the center of mass, respectively. It is important to indicate that the total energy of system is determined according to the center of mass frame.

\section{Radial Equations}

By substituting the definitions in Eq.(\ref{components}), expressions in Eq.(\ref{aaaa1}), the matrices in Eq.(\ref{matrices}) and Eq.(\ref{potential}) into the Eq.(\ref{Eq1}), a set of equation can be obtained in terms of relative coordinates
\begin{eqnarray}
\left( \epsilon \left( r\right) -2b\right) \zeta _{1}\left(
r_{1},r_{2}\right)
&-&{\overset{\frown }{\partial }_{-}}\zeta_{2}\left( r_{1},r_{2}\right)\nonumber\\
&+&{\overset{\frown }{\partial }_{-}}
\zeta _{3}\left( r_{1},r_{2}\right) =0, \label{equation4}
\end{eqnarray}
\begin{eqnarray}
\epsilon \left( r\right) \zeta _{2}\left( r_{1},r_{2}\right)+{\overset
{\frown }{\partial }_{+}}\zeta _{1}\left( r_{1},r_{2}\right) +{
\overset{\frown }{\partial }_{-}}\zeta _{4}\left( r_{1},r_{2}\right) =0, \label{equation5}
\end{eqnarray}
\begin{eqnarray}
\epsilon \left( r\right) \zeta _{3}\left( r_{1},r_{2}\right)-{\overset{\frown }{\partial }_{+}}\zeta _{1}\left( r_{1},r_{2}\right)-{\overset{\frown }{\partial }_{-}}\zeta _{4}\left( r_{1},r_{2}\right) =0, \label{equation6}
\end{eqnarray}
\begin{eqnarray}
\left( \epsilon \left( r\right) +2b\right) \zeta _{1}\left(
r_{1},r_{2}\right)&-&{\overset{\frown }{\partial }_{+}}\zeta
_{2}\left( r_{1},r_{2}\right)\nonumber\\
&+&{\overset{\frown }{\partial }_{+}}
\zeta _{3}\left( r_{1},r_{2}\right) =0,  \label{equation7}
\end{eqnarray}
where
\begin{eqnarray}
\epsilon \left( r\right)&=&\left( \frac{w}{c}+V\left( r\right) \right),\nonumber\\
b&=&\frac{m_{e}c}{\hslash },\nonumber\\
V\left(r\right)&=&-\frac{\alpha}{r},\label{epsilon}
\end{eqnarray}
and $\alpha$ is fine structure constant.\\

\qquad The space-time is transformed into the polar space-time so that we can construct all possible spin eigen-state for o-Ps and to exploit the angular symmetry of polar space-time background. We conducted this calculation mentioned in the previous sentence by using both
\begin{eqnarray*}
\overset{\frown }{\partial }_{-}=\left( \partial _{r_{1}}-i\partial _{r_{2}}\right) = e^{-i\phi }\left( -
\frac{i}{r}\partial _{\phi }+\partial _{r}\right)
\end{eqnarray*}
and
\begin{eqnarray*}
\overset{\frown }{\partial }_{+}=\left( \partial _{r_{1}}+i\partial _{r_{2}}\right)=e^{i\phi }\left( \frac{i}{
r}\partial _{\phi }+\partial _{r}\right)
\end{eqnarray*}
which standing for the annihilation and creation angular momentum operators, respectively. Now, adding and subtracting, we can write the equations in Eq.(\ref{equation4}), Eq.(\ref{equation5}), Eq.(\ref{equation6}) and Eq.(\ref{equation7}) as follows;

\begin{equation}\label{1a}
\left.\begin{aligned}
\left( \epsilon \left( r\right) -2b\right)& \zeta _{1}\left( r,\phi \right)\\
\phantom{\quad}& \phantom{a} -
2e^{-i\phi }\left( -\frac{i}{r}\partial _{\phi }+\partial
_{r}\right) \zeta _{2}\left( r,\phi \right) =0
\end{aligned}\right.
\end{equation}
\begin{eqnarray}
\epsilon \left( r\right) \zeta _{2}\left( r,\phi \right)&+&e^{i\phi }\left( \frac{i}{r}\partial _{\phi }+\partial _{r}\right) \zeta _{1}\left(
r,\phi \right)\nonumber\\
&+&e^{-i\phi }\left( -\frac{i}{r}\partial _{\phi
}+\partial _{r}\right) \zeta _{4}\left( r,\phi \right) =0,\label{1b}
\end{eqnarray}
\begin{equation}\label{1c}
\left.\begin{aligned}
\left( \epsilon \left( r\right) +2b\right) &\zeta _{1}\left( r,\phi\right)\\
\phantom{\quad}& \phantom{a} -2e^{i\phi }\left( \frac{i}{r}\partial _{\phi
}+\partial _{r}\right) \zeta _{2}\left(r,\phi\right)=0.
\end{aligned}\right.
\end{equation}
By multiplying the Eq.(\ref{1a}) with $e^{i\phi }$ and the Eq.(\ref{1c}) with $e^{-i\phi }$ and moving the exponential terms in Eq.(\ref{1b}) in front of the spinor components carefully, due to the consistence of equations, we can rewrite the equations in terms of new spinor components corresponding to the $m_{s}$ (possible values of $m_{s}$ are +1,0,-1), eigen-states as in the following equation set.
\begin{eqnarray}
\left( \epsilon \left( r\right) -2b\right) \zeta _{1}\left( r\right)
e^{i\phi }-{2}\partial _{r}\zeta _{2}\left( r\right) =0,\label{aa1}
\end{eqnarray}
\begin{eqnarray}
\epsilon \left( r\right) \zeta _{2}\left( r\right) +\partial
_{r}\zeta _{1}\left( r\right) e^{i\phi }+\partial _{r}\zeta
_{4}\left( r\right) e^{-i\phi }=0,\label{aa2}
\end{eqnarray}
\begin{eqnarray}
\left( \epsilon \left( r\right) +2b\right) \zeta _{4}\left( r\right)
e^{-i\phi }-{2}\partial _{r}\zeta _{2}\left( r\right) =0.\label{aa3}
\end{eqnarray}
By adding and subtracting of Eq.(\ref{aa1}) with Eq.(\ref{aa3}) and re-arranging of the Eq.(\ref{aa2}), with using following definitions provide simplicity
\begin{eqnarray}
\zeta _{+}(r,\phi)&=&(\zeta _{1}(r)e^{i\phi }+\zeta _{4}(r)e^{-i\phi}),\nonumber\\
\zeta _{-}(r,\phi)&=&(\zeta _{1}(r)e^{i\phi }-\zeta_{4}(r)e^{-i\phi }),\nonumber\\
\zeta _{0}(r)&=&\zeta _{2}(r).\nonumber\\
\end{eqnarray}
The equation set (Eq.(\ref{1a}), Eq.(\ref{1b}) and Eq.(\ref{1c})) become as in written in Eq.(\ref{SET})
\begin{eqnarray}
&\epsilon \left( r\right) \zeta _{+}\left( r,\phi\right) -2b\zeta _{-}\left(r,\phi\right) -{4}\partial _{r}\zeta _{0}\left( r\right) =0,\nonumber\\
&\epsilon \left( r\right) \zeta _{-}\left( r,\phi\right) -2b\zeta _{+}\left(r,\phi\right) =0,\nonumber\\
&\epsilon \left( r\right)\zeta _{0}\left( r\right) +\partial
_{r}\zeta _{+}\left(r,\phi\right) =0.\label{SET}
\end{eqnarray}

\section{ANNIHILATION ENERGY AND LIFE-TIME OF THE o-Ps}

The annihilation energy and the life-time of an o-Ps is acquired by solving following equation derived from the set Eq.(\ref{SET}):
\begin{eqnarray}
\overset{\mathbf{\cdot \cdot }}{\zeta _{+}\left( r\right) }-\frac{\overset{\mathbf{\cdot }}{\epsilon\left( r\right) }}{\epsilon \left( r\right) }
\overset{\mathbf{\cdot }}{\zeta _{+}\left( r\right) }+\frac{\epsilon \left(
r\right) ^{2}-4b^{2}}{4}\zeta _{+}\left( r\right) =0.\label{s4}
\end{eqnarray}
To simplify the solution function, following expression can be defined.
\begin{eqnarray*}
\zeta _{+}\left( r\right) =e^{\left( -\frac{r}{2c}\sqrt{4b^{2}c^{2}-w^{2}}
\right) }r^{\frac{i\alpha}{2}}\overset{\mathbf{\sim }}{\zeta _{+}\left( r\right)}.
\end{eqnarray*}
And then, for $z=-\frac{wr}{\alpha c}$ as a new dimensionless independent variable, we obtained the following differential equation;
\begin{eqnarray}
\overset{\mathbf{\cdot \cdot }}{\zeta _{+}\left( z\right) }&+&\left[ \alpha
_{1}+\frac{1+\beta }{z}+\frac{1+\gamma }{z-1}\right]\overset{\mathbf{\cdot }}{\zeta _{+}\left( z\right) }\nonumber\\
&+&\left[ \frac{\mu }{z}+\frac{\nu }{z-1}\right]
\zeta _{+}\left( z\right) =0,\label{s5}
\end{eqnarray}
where
\begin{eqnarray*}
&\alpha _{1}=\frac{\alpha}{w}\sqrt{4b^{2}c^{2}-w^{2}},\quad\beta =-i\alpha,\gamma=-2,\nonumber\\
&\delta =-\frac{\alpha^{2}}{2},\eta =1+\frac{\alpha^{2}}{2},
\end{eqnarray*}
and
\begin{eqnarray*}
&\delta =\mu +\nu -\alpha _{1}\frac{\left( \beta +\gamma +2\right) }{2},\nonumber\\
&\eta =\left( 1+\beta \right) \frac{\alpha _{1}}{2}-\frac{\left( \beta
+\gamma +\beta \gamma \right) }{2}-\mu.
\end{eqnarray*}
The solution functions of Eq.(\ref{s5}) are found to be as C-type Heun functions
\begin{eqnarray}
{\overset{\sim }\zeta _{+}\left( z\right) }&=&C_{1}HeunC(\alpha _{1},\beta
,\gamma ,\delta ,\eta ,z)\nonumber\\
&+&C_{2}z^{\left( -\beta \right) }HeunC(\alpha _{1},-\beta ,\gamma ,\delta
,\eta ,z),\label{qq}
\end{eqnarray}
providing that $C_{1}$ and $C_{2}$ are arbitrary constants, in Eq.(\ref{qq}), the second solution satisfies the boundary conditions.
To compute frequency spectrum, for Heun-C functions, following relation is used
\begin{eqnarray*}
\delta +(n+1+\frac{1}{2}(\beta +\gamma ))\alpha _{1}=0.
\end{eqnarray*}
The frequency spectrum of the system is found in terms of principal quantum number, $n$, and the fine-structure constant
\begin{eqnarray}
w_{n}=\frac{2m_{e}c^{2}}{\hslash }\sqrt{\frac{n^{4}+\frac{3}{4}n^{2}\alpha^{2}-i
\frac{\alpha^{3}n}{8}}{n^{4}+n^{2}\alpha^{2}}}.\label{fre}
\end{eqnarray}
The obtained frequency expression in Eq.(\ref{fre}) can be clarified as
\begin{eqnarray*}
w_{n}=\resizebox{0.8\hsize}{!}{$\frac{2m_{e}c^{2}}{\hslash}\left[\frac{n^{8}+\frac{3}{2}\alpha^{2}n^{6}+
\frac{9}{16}\alpha^{4}n^{4}+\alpha^{6}n^{2}/64}{n^{8}+2\alpha^{2}n^{6}+\alpha^{4}n^{4}}\right] ^{\frac{1}{4}}e^{i\frac{\theta }{2}}$},
\end{eqnarray*}
here
\begin{eqnarray*}
\theta =tan^{-1}\left[-\frac{\alpha^{3}n
}{8n^{4}+6\alpha^{2}n^{2}}\right].
\end{eqnarray*}
The principal quantum number becomes only positive real number,$\ n\geq 1$, in the solution of C-type Heun differential equation. Also, we see that the frequency expression is complex that has a real part
corresponding to proper oscillation and an imaginary part corresponding to damping oscillation of the system, and that, for the large n values the o-Ps system is in resonance state before exactly its damping, because the imaginary part goes to zero before the real part. And, applying this real part of frequency spectrum to our case, the annihilation energy of the o-Ps ($\approx 2m_{e}c^{2}-6.8$, eV) can be calculated, in the ground state where the binding energy is $-1,089985698\quad10^{-18}\quad J$. Furthermore, from the imaginary part of the frequency relation, one can end up with the instability of o-Ps (the system actually decays in a very short-time), spontaneously in our calculations.\\

\begin{table}[ht]
\centering
\begin{tabular}{ | c | c | c |}
\hline
    n     &      Imw (Hz)        &  Decay Time (s)    \\ \hline
   $1$    &     $3,776*10^{13}$  & $0.026*10^{-12}$   \\ \hline
   $2$    &     $4,720*10^{12}$  & $0.211*10^{-12}$   \\ \hline
   $3$    &     $1,398*10^{12}$  & $0.715*10^{-12}$   \\ \hline
   $4$    &     $5,900*10^{11}$  & $1.694*10^{-12}$   \\ \hline
   $5$    &     $3,021*10^{11}$  & $3.310*10^{-12}$   \\ \hline
   $6$	  &     $1,748*10^{11}$  & $5.720*10^{-12}$   \\ \hline
\end{tabular}
\caption{Life times of the o-Ps for a few n values in vacuum.%\textcolor[rgb]{1,0,0}%
}\label{table1}
\end{table}

\section{Conclusion}
Fundamental purpose of this research is to find exact annihilation energy and life-time of o-Ps which is a spin-symmetric state of the system formed by an electron - an antielectron pair. Due to the self annihilation behaviour of the system, the frequency spectra of o-Ps should contain the imaginary part, similar to exciton which formed by an electron and a hole as we found in our previous work \cite{5}. Depending on the spin polarization of the Ps, the life-times of $o-Ps$ and $p-Ps$ differ from each other (while o-Ps decays in nanoseconds, p-Ps decays picoseconds approximately). Infact, this difference (three order of magnitude) originates from the spin-spin interaction \cite{11} force. Regardless of its type ($o-Ps$ or $p-Ps$), the exact life-time can be interpreted as the duration where the electron and positron annihilates each-other without any effect such as electromagnetic field \cite{3}, photon-photon interaction in final state or screening energy since the particles interact via an attractive coulomb force, predominantly. Due to the fundamental conservation of energy law,  the masses of the particles are converted to the pure energy as high energetic $\gamma$-quantas in which the number of the quantas are determined by the well-known C-symmetry \cite{9} provide that there is a difference between those as the binding energy of o-Ps in S-state. Since the obtained spectra includes imaginary part, the life-time (actually decay time of the o-Ps) spontaneously in our model. We conclude from Eq.(\ref{fre}) that the binding energy of the system for $n=2$ is lower than the binding energy for $n=1$. Moreover, our frequency spectrum clearly indicates that the life-time of the system for $n=2$ level is longer than the life time of ground state of the system, (see table \ref{table1}).

\begin{acknowledgments}
We thank Prof. Dr. Nuri Unal, Dr. Ramazan Sahin, Dr. Ganim Gecim, Dr. Semra Gurtas Dogan and Cavit Tekincay for useful discussions.
\end{acknowledgments}

%\bibliography{apssamp}
%apsrev4-2.bst 2018-12-27 (MD) hand-edited version of apsrev4-1.bst

\end{document}